\documentclass[prb,aps,preprint]{revtex4-1}

\usepackage{amsmath}
\usepackage{amssymb}
\usepackage{amsbsy}
\usepackage{graphicx}
\usepackage{upgreek}

\newcommand{\unit}[2]%
{\mbox{\ensuremath{#1}}\mbox{\,\ensuremath{\mathrm{#2}}}}
\newcommand{\Jc}{\ensuremath{J_\mathrm{c}}}
\newcommand{\JcB}{\ensuremath{\Jc(B)}}
\newcommand{\JcH}{\ensuremath{\Jc(H)}}
\newcommand{\Bself}{\ensuremath{B_\mathrm{sf}}}

\begin{document}

\title{Numerical extension of the power law~\JcB{}
to zero~field in thin superconducting~films}

\author{F.~Hengstberger, M.~Eisterer, H.~W.~Weber}
\email{hengstb@ati.ac.at}
\affiliation{Atominstitut,
Vienna University of Technology,
Stadionallee 2,
1020 Wien}

\begin{abstract}
  Numerical simulations of the current and field distribution
  in thin superconducting films are carried out
  for a given material law \JcB{} and
  as a function of the applied field $H$,
  taking the sample's self-field into account.
  The dependence of the critical current density on the applied field 
  \JcH{} is computed for comparison with experiment,
  considering the geometry of transport measurements.

  We show
  that extrapolating the high field power law
  $\Jc \propto B^{-\alpha}$
  to the lowest fields results in a finite critical current
  at zero applied field $\Jc(H = 0)$,
  despite the singularity of \JcB{}.
  Moreover,
  particular features of the experiment,
  such as a low field plateau in \JcH,
  are reproduced
  and found to be determined by the self-field.
\end{abstract}

\maketitle

\section{Introduction}
When measuring the critical current density \Jc{} in YBCO thin films
as a function of the applied field $H$
two distinct regimes are usually observed:
starting from zero applied field,
\JcH{} first remains approximately constant
and then crosses over to a power law 
$\JcH \propto H^{-\alpha}$
dependence.
This observation is frequently discussed in terms
of a transition from one pinning regime to another.

However,
in order to extract information on pinning from,
e.g.,
a transport measurement of \Jc,
it is essential to determine
the critical current density as a function of the magnetic \emph{induction} \JcB.
Consequently,
also the contribution of the self-field \Bself{},
which stems from the supercurrents in the sample,
to $B=\mu_0H+\Bself$ has to be taken into account.
This is particularly important at low applied fields,
i.e.,
when $\mu_0H$ is comparable to or even smaller than \Bself,
and it is a priori not clear if a measurement of \JcH{}
reveals the intrinsic \JcB{} of the material.

\section{Calculation}
Numerical calculations similar to \cite{Ros07} 
allow establishing a relation between an arbitrary material law \JcB{},
which controls the current density distribution inside the sample,
and the measured quantity \JcH{},
i.e.,
the average current density at a certain applied field.
Here,
the tape cross-section
is divided into discrete elements
and initally a current density $\Jc(B=\mu_0H)$
is assigned to each of them.
After calculating the self-field distribution inside the sample
the elements are updated according to $\Jc(B=\mu_0H+\Bself)$.
The procedure is iterated until a current density
and field distribution is found,
where all the elements satisfy \JcB{},
and the average current density \JcH{} is computed.
Varying $H$ results in a \JcH{} curve,
which depends on the dimensions of the sample
and on the material law,
and enables a comparison between \JcH{} and
\JcB{} for a sample with a specific cross-section
($\unit{5}{mm}\times \unit{1}{\mu m}$ in all the simulations).

\section{Results}

\begin{figure}
  \includegraphics[width=.5\textwidth]{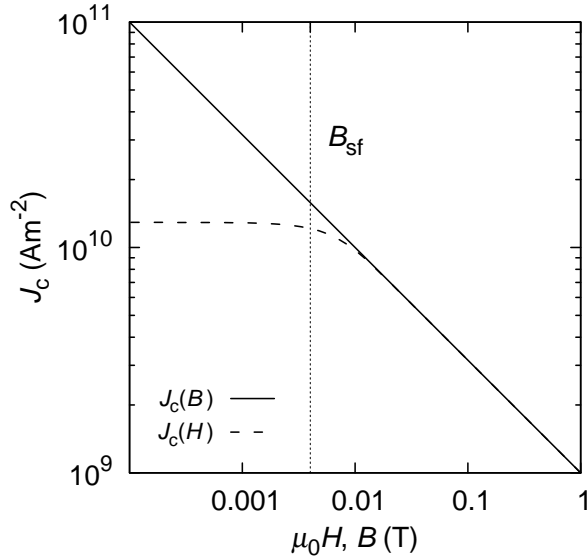}
  \caption{\label{fig:jchb}
    At high fields
    (\unit{> 0.01}{T})
    \JcH{} coincides with the material law \JcB{}
    because the self-field of the transport current is negligible.
    If the applied field decreases
    to the magnitude of the sample's self-field,
    \JcH{} is approximately constant and differs from \JcB{}.}
\end{figure}

Figure~\ref{fig:jchb} shows curves generated by
$\Jc(B) = J_1B^{-\alpha}$
to analyse the significance of the power law \JcH{}
found in transport measurements on YBCO thin films
(the parameters are $J_1=\unit{10^9}{Am^{-2}}$ and $\alpha=0.5$).
At applied fields above
\unit{0.01}{T}
\JcH{} coincides with \JcB{},
which demonstrates that in this field range a transport measurement
reveals significant information on pinning.
If,
however,
the applied field decreases to below 
\unit{0.01}{T},
\JcH{} deviates from \JcB{}
and remains approximately constant
despite the strong \JcB{} dependence.
It follows that a transport measurement of \JcH{}
does not disclose the intrinsic \JcB{} of the material in this field range.

Note,
that the extrapolation of the high-field power law \JcB{}
to the lowest fields reproduces the field independent \JcH{}
observed in experiment without any additional assumptions.
Further,
$\Jc(H=0)$ is finite regardless of the (unphysical)
divergence of \JcB{} at zero induction.%
\footnote{For stability reasons \JcB{} is cut off during the iteration,
but the final configuration satisfies \JcB{} in the entire sample.
For symmetry reasons it is necessary that
either the number of rows or columns of the discretisation is even.}
This is a consequence of the self-field:
the transport current always maintains
a certain magnetic induction inside the sample.
As long as $\mu_0H$ is negligible compared to \Bself{},
the effect on the mean transport current density is insignificant
and $\JcH{}$ is approximately equal to $\Jc(H=0)$.
The behaviour changes when $\mu_0H$ becomes comparable to $\Bself{}$.
If the applied field is further increased it gradually takes over
until it governs the current distribution at high fields
and,
as a consequence,
\JcH{} coincides with \JcB{}. 

The field,
where the transition from self-field to external field controlled
current transport occurs,
can be quantified using the magnetic field scale introduced in \cite{Bra93}
to describe magnetisation and transport currents in thin films

\begin{equation}
\label{eqn:bself}
  \Bself
  \approx
  \mu_0 c\Jc/\pi\,.
\end{equation}

Inserting the thickness
$c=\unit{1}{\mu m}$
and the critical current density at zero applied field $\Jc(H=0)$
into (\ref{eqn:bself})
results in
$\Bself \approx \unit{4}{mT}$,
which falls well in between
the constant \JcH{} at low fields
and the high field power law dependence
(see figure~\ref{fig:jchb}).

\begin{figure}
  \includegraphics[width=.5\textwidth]{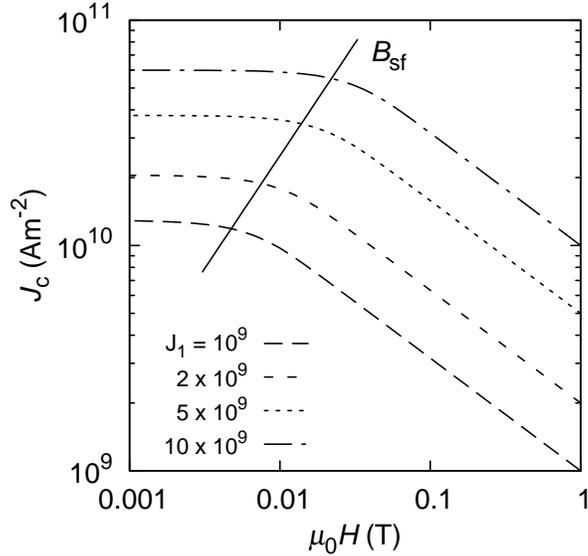}
  \caption{\label{fig:jct}
    Simulated \JcH{} curves generated by a power law \JcB{}
    with increasingly higher current densities.
    The transition into the self-field regime
    is correctly described by a simple equation.}
\end{figure}

It is clear from the above equation
that the transition between the two \JcH{} regimes
is shifted to higher fields,
when the sample supports higher current densities,
an effect frequently observed
in temperature dependent \JcH{} measurements.
Such an experiment is simulated in figure~\ref{fig:jct},
which shows a set of \JcH{} curves computed
using increasingly larger values of $J_1$
to simulate the effect of lower temperatures.
The transition into the self-field regime,
where \JcH{} becomes constant and differs from \JcB{},
is successfully described by (\ref{eqn:bself}).
The fact that the sample thickness is the only free parameter
in this equation provides an easy way to exclude self-field effects
before discussing pinning.

\section{Conclusion}
The relation between the measured \JcH{} dependence
assessed in a transport experiment
and the intrinsic \JcB{} of the material
was analysed by numerical calculations.
The computations show that \JcH{} and \JcB{} significantly differ,
if the applied field is comparable to or smaller than
the field generated by the transport current in the sample.
If the applied field is in this range,
\JcH{} remains approximately constant
and does not reveal information on \JcB{} at the same field.
An expression,
which depends only on the thickness of the sample,
allows to ensure that the observed \JcH{}
reflects a material property and is not a self-field effect.

\end{document}